# A Knowledge-based Treatment of Human-Automation Systems

Yoram Moses and Marcia K. Shamo

*Abstract*—In a supervisory control system the human agent's knowledge of past, current, and future system behavior is critical for system performance. Being able to reason about that knowledge in a precise and structured manner is central to effective system design. In this paper we introduce the application of a well-established formal approach to reasoning about knowledge to the modeling and analysis of complex human-automation systems. An intuitive notion of knowledge in human-automation systems is sketched and then cast as a formal model. We present a case study in which the approach is used to model and reason about a classic problem from the human-automation systems literature; the results of our analysis provide evidence for the validity and value of reasoning about complex systems in terms of the knowledge of the system's agents. To conclude, we discuss research directions that will extend this approach, and note several systems in the aviation and human-robot team domains that are of particular interest.

## I. INTRODUCTION

A human operator's knowledge regarding a complex system's past, current, and future behavior is elemental to system performance. Indeed, issues of the existence, content, and validity of operator knowledge in a complex human-automation system underlie many major threads of research in the human-machine systems community. These research threads include broad and active domains such as *situation awareness* (briefly, an operator's knowledge of the relevant elements in the environment), *mode awareness* (operator knowledge of a system's current operational state), and *mental models* (knowledge and beliefs regarding a system's components, possible behaviors, and interdependencies). In a basic sense, all these domains are concerned with aspects of the operator's knowledge and aim to define both theoretical frameworks and evaluative criteria for determining whether that knowledge is satisfactory for system performance.

Since human operator knowledge is a fundamental element of system performance, we argue that it is important to reason about that knowledge, and its role in controlling system performance, in a precise and structured manner. Moreover, as increasingly complex human-automation and human-robot systems are developed, the need for tools that support the design and evaluation of these systems from the perspective of the agents' knowledge becomes an imperative. Consider, for instance, the knowledge possessed by the human and automation agents in advanced flight deck information systems, in systems of human supervisory controllers and multiple autonomous vehicles, or in robot-assisted search and rescue teams. Sophisticated tools are required to effectively analyze the complex and subtle interactions between what the human and non-human agents know in these and similar systems.

To be useful, a reasoning approach must satisfy a number of modeling and analysis requirements. In particular, knowledge in a system must be a well-defined entity or resource that can be directly represented, manipulated, and analyzed. Properties of knowledge unique to human agents must be expressible. Since our focus is the correct performance of the complete system, and since non-human agents play equally important albeit different roles, we will in many cases need to be able to capture and reason about the 'knowledge' of the non-human agents in the system. Finally, since systems are dynamic, it will be necessary to represent the evolution of knowledge over time, and how that knowledge both influences and is influenced by system behavior.

In this paper we introduce a formal approach to reasoning about knowledge in human-automation systems that is based on a model of knowledge developed by Halpern and Moses and their colleagues [1]-[3]. Their methodology combines the formal rigor of epistemic logic with intuitive and direct notions of knowledge and action in multi-agent systems, and thus provides a means of reasoning about systems at the level of the agents' knowledge. To date, existing applications of the approach focus primarily on systems in which all agents are non-human (e.g., communication protocols for distributed computer systems [1], [4], robot motion planning [5], adding notions of knowledge and communication to discrete event control systems [6], [7]). We argue that this formalism can be equally valuable for reasoning about knowledge in systems in which one or more agents are human, if properties of knowledge that are unique to human agents can be expressively captured.

The value of reasoning about knowledge in the analysis of socio-technical systems has already been noted in the literature [8], [9]. In the papers cited, the authors discuss an approach combining knowledge, timed automata and Activity Theory, and propose a case study of an aviation accident. From the brief description presented, it appears that their work focuses on the temporal aspect and does not model knowledge explicitly. Our approach makes use of a fine-grained analysis of knowledge based on a careful model of the different entities' local states, along the lines of Fagin *et al.*[1]. This enables us to establish rigorous claims regarding the roles of knowledge, and the lack of knowledge, in human-automation interaction and supervisory control of complex systems. Our contribution is thus to propose a cohesive formal framework for the modeling and analysis of knowledge in these systems.

In this article, we present the theoretical foundations of our approach, and sketch a case study. The article is structured as follows. Within the domain of complex human-automation systems in which the human's role is that of supervisory controller, we first define more precisely a notion of human knowledge in a complex system. We also put forth basic criteria for what it means for a human operator's knowledge to be satisfactory for system performance. We then describe the main concepts and elements of the knowledge formalism as developed by Fagin *et al.*, and present a variation that provides a more expressive language for modeling human-automation systems. Following, we consider a case study in which our formalism is used to model and reason about the *Therac-25 device* problem, a classic problem in the human-automation systems literature [10]. We provide a brief overview of the relevant aspects of the problem, and construct a knowledge-based model. We then use the model to evaluate the knowledge in the system against formally defined criteria, and show how a solution can be derived and then validated. To conclude, we present our views regarding the potential value of the approach for reasoning about knowledge in a variety of human-automation and human-robot systems, and outline a number of research directions that will expand the value of the approach as a means of modeling complex systems.

## II. HUMAN KNOWLEDGE IN COMPLEX SYSTEMS

Human knowledge is a significant and intriguing topic for study in many domains. Philosophers from ancient times have considered fundamental questions such as the nature of knowledge and the origin of its existence, and began the development of logics and formal modeling in order to more precisely reason about these complex and often subtle concepts [11]. In more contemporary times, scholars in fields such as Artificial Intelligence and Cognitive Science have variously interpreted and re-interpreted the concept of knowledge in order to explore their particular domains, focusing for example on notions such as knowledge representation and knowledge-based agents that are fundamental to AI. Valuable outgrowths of work in these and other important fields include both strong support for the critical value of knowledge as a formal construct, and richly descriptive tools for precisely describing and reasoning about knowledge [12]-[15].

Our own goal is to draw on this formal approach as a means of reasoning precisely about complex human-automation systems. More specifically, we aim to use this well-structured notion of knowledge to model and analyze the design of complex systems for which the human's knowledge is required for supervisory control. The ability to describe in exact terms what a human operator must know in order to control a system, coupled with the ability to describe, in like terms, what that human can be expected to know as a function of the interface, offers a rigorous approach to the analysis of complex human-automation system design efficacy.

### A. A Notion of Operator Knowledge

So that we may reason formally about operator knowledge in the context of complex systems, we first need to conceptualize operator knowledge in supervisory control in a manner that will be amenable to capture within a formal framework. In this section we discuss *types* of knowledge that an operator may need to control a complex system, and then examine several additional characteristics of operator knowledge that should also be captured in our framework.

### 1) Knowledge Types

The concept of knowledge in the literature is often partitioned and variously classified, with each research domain categorizing knowledge in the way that is most suited for its theories and constructs. Since our focus at present is the knowledge of a human supervisory controller within the context of a complex system, we need to cast our notion of knowledge accordingly. What is the knowledge that is required for supervisory control?

Sheridan notes that to function as a supervisory controller, the human agent must be able to command the system in accordance with defined specifications, monitor the system's behavior to ensure that it performs as required, and identify and correct anomalous system behaviors [16], [17]. It is straightforward to interpret these supervisory control tasks in terms of knowledge: the effective human operator needs both knowledge of the current state and behaviors of the system, and knowledge of the rules defined by the system specifications. To represent these knowledge requirements correctly, our approach will need to capture the knowledge of current system state available from the automated system's interface (e.g. the displays) as well as the knowledge that the human operator is expected to possess regarding the rules of possible and acceptable system behavior.

a) *Explicit knowledge*

The information available from the human-automation interface at any given point in time is the source of the operator's *explicit* knowledge.[1] We wish to say, for example, that a competent human driver explicitly knows that the current speed of the car she is driving is 40 miles per hour since that value is displayed on the speedometer. While the relationship between displayed information and useful knowledge may depend on non-trivial factors such as the agent correctly perceiving the display and understanding its meaning, the compatibility of the display format to the task, and so forth [18], we assume here that the display type, quality, and information saliency are adequate for the task at hand. Explicit knowledge is needed for supervisory control tasks such as monitoring system health and status; it is also necessary for tasks such as synchronizing the actions of interacting or otherwise interdependent system components (e.g., not opening a tank cover until the internal pressure of the tank has dropped below some pre-defined level).

b) *Mental model knowledge*

A second type of knowledge needed for human-supervisory control is the 'knowledge-in-the-head' that the human agent possesses regarding the global behaviors and properties of the physical system with which he is interacting [19]-[21]. This *mental model* knowledge can be thought of as a collection of inferences that the operator knows to make about the system. Mental model knowledge is required both for reasoning about and interpreting current interface information as well as for future-oriented supervisory control tasks such as planning and scheduling, troubleshooting, and decision-making [16]-[18]. We assume that the operator's mental model knowledge also includes what we shall call *ground knowledge* – essential domain-relevant knowledge that the human agent may be assumed to possess, such as basic logical and computational skills. We suggest that this knowledge is 'automatic' in nature and thus requires almost no effortful thinking or reasoning action [18], [22]. For instance, it seems correct to assume that a human agent who is sufficiently skilled to act as supervisory controller of a complex system, should be able to automatically compare two displayed integers and determine which of them is greater.

Mental model knowledge is normally based in fact (e.g. through training). However, since it evolves over time as a result of repeated interactions, mental model knowledge may grow to include some proportion of unproven (and perhaps incorrect) beliefs [23]. For any typically-sized complex system the human operator's mental model knowledge is clearly incomplete, as the number of potential system behaviors under all possible conditions is too large to be known.

[1] It is worth mentioning that the human-automation interface need not be visual; our intent is any interface that provides the operator with information relevant to system behavior.

### 2) Characteristics of Human Knowledge

In order to develop a representation that allows us to reason accurately as well as expressively about human supervisory control knowledge within the context of complex systems, we need to consider several additional aspects or characteristics of operator knowledge.

#### a) Reasoning and knowledge

In a practical sense, what the operator knows at any given moment is the result of interpreting the explicit knowledge available using the mental model knowledge. Precisely how humans reason, the rules of inference that guide their reasoning activities, or indeed whether humans use any form of mental logic at all are on-going and strongly debated questions in the literature [24]. The goal of this paper is not to add to that debate, nor to make claims regarding the actual processes by which humans reason, or the specific rules of logical inference they may use. To capture a notion of reasoning useful for our purposes, we will assume that the human supervisory controller has the ability to reach at least a particular set of conclusions that are generated by a small set of deductions that are part of her mental model.

For example, consider that at a given moment in time the driver agent explicitly knows that the allowable speed is 25 mph (she just passed the speed limit sign), and that her current actual speed is 40 mph as displayed on the speedometer. Further, assume that the set of facts that comprise her mental model includes the inference that an actual speed greater than an allowable speed may result in a speeding ticket. The inference will then allow the human agent to deduce the (quite important) conclusion that she is currently in danger of receiving a ticket.

#### b) Human computational ability

A related characteristic of human reasoning ability and the knowledge that results is the question of the operator's *computational power* – even if the operator only reaches conclusions based on the inferences in her mental model, how many conclusions can she reach in a bounded period of time? If a conclusion requires conclusions from other inferences as inputs or antecedents, how long can this chain of inferences be before the human operator is overwhelmed? As research shows, human computational power is quite limited [18], [22], [25] and even more so in the time- or safety-critical environments within which most complex systems function [26]. In order to capture some reasonable facsimile of the knowledge of a supervisory controller, our approach will have to limit the number of inferences that the operator can make at a given moment.

## B. When Does An Operator Know Enough?

If our aim is to provide a practically useful methodology and tool set for the design and analysis of complex systems from the perspective of agent knowledge, we must provide not only a formal means of describing knowledge in these systems, but also formal metrics against which knowledge in systems can be rigorously evaluated. Though the human agent's knowledge of a complex system's possible behavior is incomplete by definition due to the size of the state space, we propose that at a minimum, the human agent's knowledge must be *sound* and *adequate*. We describe these properties next, and note that if they hold, we shall say that the human agent's knowledge is *satisfactory*.

### 1) Soundness

We will say that the human agent's knowledge is *sound* at a given system state if the facts that the agent is said to know and the inferences that the agent can make there are true. As we discuss in following sections, soundness is a fundamental property of knowledge that is formalized as the *Knowledge Axiom*, stating that only what is true can be known [1]. A formal definition of knowledge for human agents is sound if the Knowledge Axiom holds at all states of the system of interest.

### 2) Adequacy

While an agent's knowledge in a complex system will necessarily be incomplete, the knowledge must be *adequate*; in particular, it must guarantee her ability to identify anomalous or 'illegal' behaviors or states (i.e. not in accordance with system specifications). In other words, though the agent will not know, *a priori*, all the possible behaviors that a system may exhibit, she must always be able to determine when the current system behavior is 'bad'.

Having discussed basic notions related to human knowledge in supervisory control, we next provide a brief introduction to the basic elements of the knowledge formalism in order to recast our model.

## III. REASONING ABOUT KNOWLEDGE – AN INTRODUCTION

### A. Elements of the Framework

The knowledge formalism was originally adapted to distributed and multi-agent systems by Halpern and Moses and was expanded on in numerous writings [1], [2], [5], [27]-[30]. It includes structures to represent the entities or agents in the system and the knowledge they can be said to possess. The formalism also includes a means for describing the interaction of knowledge and action and the behavior of the system. It thus provides a rigorous methodology for reasoning about knowledge in a dynamic multi-agent system, and for analyzing and proving properties of that knowledge. Here we focus on the elements of the formalism that are relevant to the present work, and describe how knowledge is defined using these elements. The primary references for this part are [1]-[3]. Once the foundations of the formalism have been presented, we introduce a number of concepts that will enable more expressive representation of knowledge properties of human agents.

*1) Agents*

The formalism considers *agents* and systems of multiple interacting agents, where an agent might be a robot, a processor, a human, a physical object, or any other entity of interest in the system. The approach includes the external environment as an agent, albeit a special type of agent, whose behavior is not under the control of the other agents in the system. In general, the environment's role is to represent all that is relevant to the system that is not captured by the other system agents. In this paper we focus on systems involving a single human agent $h$, (intuitively, this is the operator), a single "automaton" agent $a$ – the complex system that $h$ operates, and the environment agent $e$. In modeling the speedy driver, for example, we could capture the driver as human agent $h$, the car as the automaton agent $a$, and the outside world (including, for instance, traffic lights and whether or not there is a policeman watching) as the environment $e$.

*2) Local states, global states, and agent knowledge*

When reasoning about knowledge, we are typically interested in modeling a dynamic situation in which the world evolves, and with it the state of knowledge of the agents. In this section we consider a natural way to model these.

We think of every agent at any given instant as being in a well-defined local state. The local state in our setting captures all of the information that is available to the agent at a given instant. This is the information available to the agent when it determines its next action. The precise structure and contents of this local state depends on the system that is being modeled.

A *global state* corresponds to a snapshot of the state of the agents in the world frozen at an instant. In our three-agent setting, a global state would be modeled by a triple ($s_e$, $s_h$, $s_a$), where $s_i$ is agent $i$'s local state, for $i=e,h,a$. Intuitively, an agent's local state captures exactly what is visible to the agent at the current point - the agent's knowledge depends solely on its local state. The agent is able to distinguish between two global states exactly if its local state in one is different from its local state in the other. For example, consider a car in which there is an indicator light for low brake fluid. As long as the indicator light is off, the driver cannot distinguish whether or not the brake fluid is leaking. All the driver knows is that the fluid level is not below the critical level. Indeed, if the indicator is not fully reliable, then the driver cannot distinguish a situation in which the indicator is working and the fluid is sufficient, from one in which the fluid level is critical but the indicator is malfunctioning. The formalism thus allows us to accurately model and reason about situations in which agents have only partial knowledge of system behavior. The state $s_e$ is called the local state of the environment, and it accounts for all else that is relevant to the analysis, possibly keeping track of aspects of the world that are not part of any agent's local state (e.g., a traffic light).

### 3) Dynamic system behavior

The evolution of a world over time produces a history, which in our terminology will be called a *run*. Formally, a run $r$ is a function from time to global states, assigning to every time instant $m$ a global state $r(m)$. If $r(m) = (s_e, s_h, s_a)$, then we denote by $r_i(m)$ the local state $s_i$, for $i=h,a,e$.

It is often convenient to identify time with the natural numbers, in which case the run is identified with the sequence $r = r(0), r(1), \ldots$. Facts, represented in our framework by formulas of $\mathcal{L}$, are considered true or false at a *point (r,m)*, consisting of a run $r$ at a time $m$. The truth of non-epistemic facts (ones that do not involve knowledge) at *(r,m)* can usually be determined based on the global state $r(m)$ at that point.

We choose to model the types of human-automation systems that we are interested in as consisting of three agents; the human agent $h$, the automation agent $a$, and the environment agent $e$. With this construction we can focus on the human 'knower' $h$, and capture how $h$'s knowledge influences and is influenced by the behavior of the complete system. Thus, at any given time each agent $i$, for $i = e, h, a$, will be in a local state $l_i$, representing all the information that $i$ has available at that time. We denote by $L_i$ the set of $i$'s possible local states. As in the general case noted above, the tuple of all the agents' local states is a global state $g$. The set $G$ of all global states is then the Cartesian product $G = L_e \times L_h \times L_a$ of the sets of local states. In a given application, there can be dependencies between elements in the agents' local states (e.g. shared data) and the set of global states that will appear in a relevant set of runs $R$ will usually be a subset of $G$.

The dynamic behavior of the system is represented by the set of runs $R$ is generated by the joint actions of the environment, automation, and human agents.

### 4)  *The logical language*

We need a language in order to rigorously reason about knowledge in complex systems, and so now formally define a logical language for this purpose. In any given application, there are a (typically small) number of basic facts of interest. These can be the status of the traffic light, the speed of the car, the direction it is driven at, etc. We call such basic facts *primitive propositions*. In each application or system being considered, we assume that a set $\Phi$ of the relevant primitive propositions is given. These will be the basic formulas of our logical language. We create more complex formulas inductively by applying logical connectives to simple formulas. Formally, we will be working with a logical language $\mathcal{L}$, which is the set of formulas defined as follows

- Every primitive proposition $p \in \Phi$ is a formula,

- If $\varphi$ and $\psi$ are formulas then so are

- $\neg\varphi$         (standing for ***not*** $\varphi$)

- $\varphi \wedge \psi$    (standing for $\varphi$ ***and*** $\psi$),  and

- $K_i\varphi$        (standing for *agent $i \in \{h,a\}$* ***knows*** $\varphi$).

We define only the two Boolean operators $\neg$ and $\wedge$ because all other Boolean operators are definable using these two [31]. (For example, $\varphi \Rightarrow \psi$   corresponding to "$\varphi$ *implies* $\psi$", can be defined as shorthand for $\neg(\varphi \wedge \neg\psi)$.) The language $\mathcal{L}$ allows us to express rich and complex statements such as $\varphi \wedge \neg K_h\varphi \Rightarrow \psi$ which reads that "if $\varphi$ is true and agent $h$ does not know $\varphi$, then $\psi$ must be the case." Such a formula would be true, for example, if $\varphi$ stands for the brake fluid level being critical, and $\psi$ stands for the fact that the indicator is malfunctioning. In fact, while we will not expand into a discussion of one agent's knowledge of another agent's knowledge in this paper, our language allows us to talk about what the human knows about the automaton's knowledge, what the automaton knows about the human's knowledge, etc.

The propositions $p,q,... \in \Phi$ and indeed all formulas of $\mathcal{L}$ are initially strings that we may intend to ascribe meaning to. We consider that the set of propositions $\Phi$ represents the basic facts about the system that may be derived from the design specification. The truth of a primitive proposition $p \in \Phi$ is defined by way of an *interpretation* π that maps every φ and *g* to True or False. Thus, if π(*p*)(*g*)=True then proposition *p* is true at the global state *g*.

Knowledge changes over time as an agent learns new facts and forgets some that are known. Thus a formula is considered true or false at a time *m* in a history (or run) *r* in a system or model *R*. We can think of a world as consisting of a tuple (*R*,π,*r*,*m*) where *R* is a system, π is the interpretation for the propositions of Φ at states of *R*, *r* is a history (or run) in *R*, and *m* is the point in time at which we are evaluating truth of formulas. The pair (*r*,*m*) is a point, and if *r* is in *R*, then it is a point of *R*. We denote the fact that a formula φ is true, or *satisfied* at a world (*R*,π,*r*,*m*) by (*R*,π,*r*,*m*) |= φ . The satisfaction relation |= is formally defined by induction on the structure of φ.

Primitive propositions $p \in \Phi$ form the base of the induction, and their truth is determined according to the interpretation π:

$$(R,\pi,r,m) \models p \text{ (for a primitive proposition } p \in \Phi) \text{ iff } \pi(p)(r(m)) = \text{True} . \quad (1)$$

Negations and conjunctions are handled in the standard way:

$$(R,\pi,r,m) \models \neg\psi \text{ iff } (R,\pi,r,m) \models \psi. \quad (2)$$

$$(R,\pi,r,m) \models \psi \wedge \psi' \text{ iff both } (R,\pi,r,m) \models \psi \text{ and } (R,\pi,r,m) \models \psi'. \quad (3)$$

For a formula $K_i\psi$ , the clause $(R,\pi,r,m) \models K_i\psi$ states that $K_i\psi$ (*i* knows $\psi$) is satisfied at the point (*r*,*m*) in system *R*. We will now discuss when this may hold.

*5) Knowledge as truth in all possible worlds*

The formal definition of knowledge is traditionally framed within the notion of possible worlds [13], [14]. Two points *(r,m)* and *(r',m')* are considered to be indistinguishable for agent *i*, which we denote by *(r,m)~$_i$(r',m')*, if *i* has the same local state in the global state *r*(*m*) as in *r'*(*m'*).

Intuitively, if *(r,m)~$_i$(r',m')* then at *(r,m)* agent *i* can equally imagine that the actual state is *(r',m')*. Hence *(r',m')* is considered a possible world for *i* at *(r,m)*.

The possible-worlds definition of knowledge states that an agent knows a fact precisely if this fact is true in every state or world that the agent considers possible. If two points *(r,m)* and *(r',m')* are indistinguishable to an agent, and the fact φ is false at *(r',m')*, then at *(r,m)* the agent cannot be said to know that φ is true. More formally,

$$(R,\pi,r,m) \models K_i\varphi \text{ holds iff } (R,\pi,r',m') \models \varphi \text{ holds for all points } (r',m') \text{ with } r' \in R \quad (4)$$

that satisfy $(r,m) \sim_i (r',m')$.

Observe that the four clauses (1)–(4) above defining the satisfaction relation `$|=$' enable us to determine, for every formula $\varphi \in L$ and every world $(R,\pi,r,m)$, whether $\varphi$ is or is not satisfied at the world.

This definition of knowledge is a powerful and useful tool for the description and analysis of properties of multi-agent systems, and enables formal proof of agent knowledge or, alternately, the impossibility of that knowledge. However, several fundamental properties of the approach make it problematic for modeling human agents in human-automation systems. We briefly mention these properties here prior to discussing the modifications to the definition that will allow a more appropriate and expressive representation of the knowledge of human agents (cf.[13]).

6) *Logical omniscience*

One of the most important difficulties with applying the traditional possible-worlds definition of knowledge to human agents is that it implies a notion of *logical omniscience*. Intuitively, logical omniscience refers to the fact that if an agent knows a set of facts it will also know all logical consequences of that set. In our context, logical omniscience suggests that in any state an agent will know all the facts that follow from the information available in that state. That is, the agent's knowledge in this case is closed under logical inference. Though this may be a fair assumption for an agent with a powerful capacity for reasoning, the typical human agent clearly does not fall into this category.[2] The knowledge that can be ascribed to a human agent using the traditional model is thus far greater than what the agent can actually be expected to know. In order to better fit the task of analyzing systems with human agents, our modifications must provide a more realistic measure of the human agent's limited ability to infer and to reason, and the limitations on human knowledge that result.

7) *Representing false knowledge*

As we have seen, the traditional possible-worlds definition of knowledge entails the Knowledge Axiom, stating that an agent knows only true facts, $K_i\varphi \Rightarrow \varphi$ . In the case of an analysis of automated agents for whom knowledge is in any event an externally ascribed notion this does not appear to be troublesome. For human agents, however, false knowledge is a common state of affairs and a particularly relevant issue in human-automation interaction – we would wish to capture, for instance, aspects of the interface design that mislead the human operator into holding false beliefs.

---

[2] This point is most succinctly emphasized by Konolige [37] who notes that if humans were capable of unlimited reasoning, a chess player would know the outcome of a chess match immediately following the first move.

IV.  DEFINING BOUNDED KNOWLEDGE IN A COMPLEX HUMAN-AUTOMATION SYSTEM

*A.  An Appropriate 3-Tiered Syntactic Model of Human Knowledge*

We now propose a formalism in which the assumptions about the computational aspects and contents of an agent's knowledge are limited, in order to more appropriately represent the bounded character of human knowledge. Rather than using possible-worlds semantics as the primary component, we will present a syntactic model in which the known facts will be obtained based on a restricted amount of reasoning, and in which what is *boundedly* known may not be true (for related approaches to syntactic knowledge see [13], [32]).

We model the human's knowledge as a three-tiered structure.

*Tier 1 - Explicit knowledge*

At the base is the agent's *explicit* knowledge, which we intuitively think of as the raw information available from the system interface: clock readings, indicators, aural warnings, and so forth.

*Tier 2 - Automatic knowledge*

Based on the agent's explicit knowledge, we allow a set of *ground* (or "automatic") conclusions, in which various propositions that may not be explicitly represented in the state are immediately observed. For example, the local state may contain two 5-digit numbers, and the agent may be expected to know immediately which of them is larger. Similarly, a driver in a residential zone that sees the odometer reading 70 mph (as part of his explicit knowledge) would automatically know that "I am speeding" is true.

*Tier 3 – Bounded Knowledge via deductions*

On top of the explicit knowledge and the ground conclusions it affords, we think of the agent as being able to perform a finite set of *deductions*. Each of these deductions will use the fact that a set of propositions appear at the ground and explicit knowledge tiers, to conclude that the agent knows a formula of interest. This is a strictly limited process however, in that knowledge that is deduced at this level cannot be used for further rounds of deduction.

We now present the formal definition of this syntactic framework. First, we define the logical language $\mathcal{L}^+$, extending the language $\mathcal{L}$ by adding a "bounded knowledge" operator $\hat{K}_h$ :

- All formulas of $\mathcal{L}$ are formulas of $\mathcal{L}^+$,

- if φ is a formula of $\mathcal{L}^+$ then so is $\hat{K}_h$ φ (thus, $\mathcal{L}^+$ is closed under $\hat{K}_h$ ), and

- $\mathcal{L}^+$ is closed under $\neg$, $\wedge$, and $K_i$.

Next, if $\Phi$ is the set of propositions, we define the set $\Phi^c = \Phi \cup \{\neg p\colon p \in \Phi\}$ containing all propositions and their negations. We model the human agent's explicit knowledge by way of a function $f_h\colon L_h \to 2^{\Phi^c}$ specifying which elements of the set $\Phi^c$ the agent explicitly knows to be true at any given local state $l_h \in L_h$ of the human agent. Observe that explicit knowledge will satisfy the Knowledge Axiom only if the function $f_h$ ensures that all explicit knowledge is true.

The agent's ground knowledge, which we think of as being derived automatically from its explicit knowledge, is captured by way of a function $a_h\colon 2^{\Phi^c} \to 2^{\Phi^c}$ that, when applied to a subset $T \subseteq \Phi^c$ produces another such subset $a_h(T) = T'$. In our setting, the function $a_h$ will be applied to sets $T$ describing the agent's explicit knowledge. For ease of exposition, we think of the ground knowledge as extending the explicit knowledge, so that we formally require that $a_h$ *be monotonic, so that* $T \subseteq a_h(T)$.

Finally, we consider a *deduction phase* determined by a set $D_h$ (typically finite) of implications of the form

$$p_1 \wedge p_2 \wedge \ldots \wedge p_k \Rightarrow \hat{K}_h \varphi, \text{ where } p_i \in \Phi^c \text{ for } i=1,2,\ldots,k.$$

We denote by $d_h(T)$ the application of the implications in $D_h$ to a subset $T \subseteq \Phi^c$ of propositional formulas. Formally,

$$d_h(T) = \{ \hat{K}_h p \mid p \in T\} \cup \{ \hat{K}_h \varphi \mid (p_1 \wedge p_2 \wedge \ldots \wedge p_k \Rightarrow \hat{K}_h \varphi) \in D_h \text{ and } p_1, p_2,\ldots,p_k \in T\}.$$

In words, the agent's bounded knowledge is obtained from applying a fixed set of deductions to its explicit and its ground knowledge. The latter, in turn, is based on the explicitly available knowledge consisting, for example, of data presented on the man-machine interface in a control room. The fact that the deduction phase consists of a single step in which a finite number of deductions are applied enforces the boundedness of this type of knowledge, avoiding forms of logical omniscience.

Putting these elements together, we define the knowledge of the human agent $h$ by way of an *epistemic setup* $\Theta_h = (f_h, a_h, D_h)$. We think of the epistemic setup as encoding a 3-step process by which $h$'s structured knowledge is obtained as a function of its local state $l$:

*Step 1.* The function $f_h$ is applied to the local state to obtain the agent's explicit knowledge.

*Step 2.* The function $\boldsymbol{a}_h$ is applied to the result to account for the immediate conclusions that the agent draws automatically from its explicit knowledge.

*Step 3.* The set of inferences available in the agent's mental model, captured by the $D_h$ component, is applied once using the function $\boldsymbol{d}_h$ and generates the agent's knowledge at *l*.

Formally, applying $\Theta_h$ to a local state *l* yields the set $\Theta_h(l) = \boldsymbol{d}_h(\boldsymbol{a}_h(\boldsymbol{f}_h(l)))$ consisting of knowledge formulas $\hat{K}_h\,\varphi$ that the human agent is taken to know at local state $l_h$.

An example will help to clarify. Let $l_h$ be the local state of a human agent *h* piloting an aircraft, and let time instant *m* at *r(m)* be the final approach just prior to landing. Among the many indicators in the cockpit is an indicator for the status of the flaps (trailing edges of the wings – *full flaps* is typically the normal status for optimized landing safety and efficiency) and an indicator for the status of the landing gear, which may be up, transitioning, or down and locked.

The pilot's local state $l_h$ at any time will consist of the status of these two indicators; for instance $l_h =$ ((flaps.not.full, landing.gear.up) describes the pilot's local state when the flaps are not at full extension and the landing gear is folded up into the airplane. In this local state the set of propositions that are true – the pilot's explicit knowledge – are that the flaps are not fully extended and that the wheels are up. We'll call these propositions $p_1$ and $p_2$, respectively.

By applying the automatic ground knowledge function $\boldsymbol{a}_h$ on this set of propositions the pilot also knows that the airplane is not configured for landing, and this immediate conclusion, call it $p_3$, is added to the set of propositions that the pilot knows. For the third step in the process, we'll assume that a trained pilot's mental model includes an inference *d* that captures the rule '*if on final approach then full flaps and landing gear down and locked*'. Applying this inference to the set of propositions $p_1$, $p_2$, and $p_3$ for time instant *m* at *r(m)* = final approach generates the pilot agent's knowledge of the formula $\psi$ at local state *l* where $\psi$ is '*on final approach and aircraft not configured for final approach so go-around (return to altitude in order to safely configure aircraft for landing)*'. Thus, $\hat{K}_h\,\psi$ describes the pilot's bounded knowledge at *l*.

## B.  Knowledge of Bad System Behavior

A supervisory controller must at all times know whether the behavior of the system is within specified and acceptable bounds. In addition to system–specific knowledge formulas that will be part of the human agent's epistemic setup $\Theta_h$, the agent is expected to know, at each local state $l$ in which the current global state is not in accordance with system specifications, that there is a problem.. We define $p_{bad}$ as a proposition standing for '*the current behavior of the system is not acceptable*' where the notion of "acceptable" would depend on the application. Moreover, we require that $\hat{K}_h\, p_{bad}$ be included in the set $\Theta_h(l)$ at any state of the system that is not guaranteed to be within acceptable bounds.

In the example above, if we assume there is nothing that precludes the go-around maneuver then the current behavior of the system is acceptable and the formula $\hat{K}_h\, p_{bad}$, is not part of the pilot's knowledge at this local state $l$.

Alternately, if the pilot in this situation cannot execute the go-around maneuver then a bad state exists and $\hat{K}_h\, p_{bad}$ would be included in the pilot's local state.

At present we make the simplifying assumption that the human knows all the formulas in her epistemic setup $\Theta_h(l)$ from the instant that she is in $l$, and do not address the (more realistic) possibility that drawing even a simple set of inferences may require some time. While we can easily extend our approach to capture this distinction, the resulting definitions are somewhat cumbersome, and are beyond the scope of this paper. (See [33] for definitions of knowledge that take computational complexity into account.)

*C. The Epistemic System Model*

The mathematical model in which we can both define truth of formulas and generate the human agent's knowledge will be an *epistemic system* $E = (R, \pi, \Theta_h)$, where $R$ is the set of possible runs, $\pi$ is the interpretation of primitive propositions at the global states of $R$, and $\Theta_h$ is the epistemic setup of the human agent as described above. Truth of formulas in $\mathcal{L}^+$ can now be defined with respect to $E = (R, \pi, \Theta_h)$. At a given time $m$ in a run $r$, we define:

$(E,r,m) \models p$ (for $p \in \Phi$)    iff $\pi(p)(r(m)) = $ True

$(E,r,m) \models \varphi \wedge \psi$    iff both $(E,r,m) \models \varphi$ and $(E,r,m) \models \psi$

$(E,r,m) \models \neg\varphi$    iff it is not the case that $(E,r,m) \models \varphi$

$(E,r,m) \models \hat{K}_h\, \varphi$    iff $\hat{K}_h\, \varphi \in \Theta_h(r_h(m))$

Finally, since the system $R$ and the interpretation $\pi$ are components of the epistemic system $E = (R, \pi, \Theta_h)$, we can also define the truth of standard (possible-worlds) knowledge with respect to $E$. For $i = a,h,e$:

$(E,r,m) \models K_i\varphi$ iff $(E,r',m') \models \varphi$ for all points $(r',m') \sim_i (r,m)$

The latter clause will enable us to consider possible-worlds knowledge and bounded knowledge within the same framework. Recall that $\hat{K}_h$ represents a syntactic, rather than semantic, notion of knowledge. That is, what the human agent "knows" in each local state is a set of knowledge formulas that is determined by the epistemic setup $\Theta_h$. There is no *a priori* guarantee that this knowledge is true, or even consistent.

Given an epistemic system $E$, we write $E \models \varphi$ (and say that $\varphi$ is *valid*, or *tautologically true in E*) if $(E,r,m) \models \varphi$ holds for all points $(r,m)$ of $R$.

### D. Criteria for Satisfactory Bounded Knowledge

We now formalize the soundness and adequacy properties to enable the analysis of whether or not the human agent's bounded knowledge in an epistemic system $E = (R, \pi, \Theta_h)$ is satisfactory for supervisory control.

#### 1) Soundness

We say that an epistemic system $E = (R, \pi, \Theta_h)$ is *sound* if the following two statements are true:

1. For every $(r, m)$ with $r \in R$, if $q \in a_h(f_h(r_h(m)))$ then $(E,r,m) \models q$. In words, for every point $(r,m)$ in a run of $R$, if $q$ is in the human agent's explicit or automatic knowledge, then $q$ is true at $(r,m)$ in the epistemic system $E$. Moreover,

2. $E \models (p_1 \wedge p_2 \wedge \ldots \wedge p_k) \Rightarrow \varphi$ holds for every implication $p_1 \wedge \ldots \wedge p_k \Rightarrow \hat{K}_h \varphi$ in $D_h$.

Intuitively, soundness guarantees that $E \models \hat{K}_h \varphi \Rightarrow \varphi$ holds, meaning that every conclusion that the human agent draws using its mental model is true at every point $(r,m)$ in the epistemic system $E$. We will prove this fact formally in Corollary 1 below.

The importance of soundness in this context is that it ensures that what the human agent is said to know according to $\hat{K}_h$ is in fact true. If the epistemic system $E$ does not satisfy soundness, so that $E \not\models \hat{K}_h \varphi \Rightarrow \varphi$, then there is one point at which there is at least one formula $\varphi$ known to the agent according to the epistemic system $E$ is false. In this case, either an element of the agent's explicit or automatic knowledge is incorrect, or an implication in $D_h$ is incorrect (possibly both).

The notion of soundness allows us to consider situations in which the knowledge available for supervisory control is incorrect. The faulty indicator light for low brake fluid mentioned previously is an example – if the indicator is malfunctioning and providing the explicit information 'brake fluid indicator light off', the driver may boundedly know that the brake fluid is sufficient while in fact it is critically low. This is in comparison with a lack of required knowledge which is addressed by the adequacy criterion we consider next.

*2) Adequacy*

Though a human operator's knowledge of a complex system is incomplete by definition, we will say that the human's knowledge is *adequate* for supervisory control only if the operator can always recognize unsafe system states and avoid anomalous system states and behaviors.[3] That is, the agent must be able to use her own local state to determine whether the current state is acceptable. For example, assume that a well-intentioned but somewhat confused cost-cutting designer concludes that since wheels are either up or down, having multiple indicators for wheel position is wasteful and that one 'wheels up' indicator should be sufficient. At any given time, the epistemic setup of the pilot of this low-cost airplane is comprised solely of $f_h(l) = p_{up}$: 'wheels are up' or its negation $\neg p_{up}$: 'not (wheels are up)', or more familiarly, 'wheels are not up'. Landing of course is allowed only if wheels have completed transitioning and are locked in the 'down' position. However, the pilot's display does not allow her distinguish between the wheels being in transition, $\neg p_{up}$, and the wheels being locked in the down position and so the information available is not adequate for her to determine whether indeed the aircraft is configured for landing.

To formalize adequacy in epistemic systems, recall that we defined $p_{bad}$ as a proposition standing for 'the current state of the system is not acceptable.' Moreover, we required that $\hat{K}_h \, p_{bad}$ hold at all points at which the global state is not in accordance with the system specification. We then say that an epistemic system $E$ is *adequate* if the following hold:

1.  $\hat{K}_h \, p_{bad} \in \Theta_h((r_h(m))$ for all states of the system such that $(E,r,m) \models p_{bad}$, thus $p_{bad} \Rightarrow \hat{K}_h \, p_{bad}$ is valid in $E$.

Soundness must also hold so that the operator's bounded knowledge of a bad state is true:

2.  $E$ is sound, so that, in particular, $E \models \hat{K}_h \, p_{bad} \Rightarrow p_{bad}$

*E. Human Knowledge and Possible Worlds Knowledge*

Our syntactic representation of bounded human knowledge via an epistemic system $E$ makes no assumptions regarding logical omniscience, consistency of reasoning, or the actual truth value of the human agent's knowledge. By using this approach we are able to capture an expressive and reasonable description of a system with human agents while at the same time we avoid the problems inherent in the application of the possible-worlds definition. However, is there a price to be paid for this useful tool set? Is it less formal, less precise, or does it limit the conclusions that can be reached regarding what the human agent knows?

---

[3] We refer here to significantly anomalous states and behaviors that impact overall performance or safety, and not the transient anomalies that are automatically corrected and are part of any complex system.

While additional work is needed to fully determine the limitations of our framework, we do claim that our approach precludes any arbitrary or specious claims to agent knowledge. As the following theorem and corollaries show, any proposition $\varphi$ that is known by human agent $h$ in a sound system $E$ is also known when knowledge is defined as truth in all possible worlds:

**Theorem 1**: Let $E = (R, \pi, \Theta_h)$ be an epistemic system and let $\varphi \in \mathcal{L}^+$. If $E$ is sound, then $E \models \hat{K}_h \varphi \Rightarrow K_h \varphi$.

In words, this theorem states that any proposition $\varphi$ that the human agent (boundedly) knows in a sound epistemic system $E$ is also known by that agent in the possible-worlds sense.

**Proof:** Suppose that $E = (R, \pi, \Theta_h)$ is sound. We need to show that if $(E, r, m) \models \hat{K}_h \varphi$ then $(E, r, m) \models K_h \varphi$ holds as well. By definition, $(E, r, m) \models K_h \varphi$ holds if $(E, r', m') \models \varphi$ holds at all points $(r', m')$ satisfying $(r', m') \sim_h (r, m)$. So let $(E, r, m) \models \hat{K}_h \varphi$ and $(r', m') \sim_h (r, m)$. Denoting $l = r_h(m)$, we have that $r'_h(m') = l$ as well. By definition of $(E, r, m) \models \hat{K}_h \varphi$ we have that $\hat{K}_h \varphi \in \Theta_h(l) = d_h(a_h(f_h(l)))$. By definition of $\Theta_h$ and $d_h$ this means that either (a) $\varphi = p \in a_h(f_h(l))$ or (b) there is an implication $p_1 \wedge \ldots \wedge p_k \Rightarrow \hat{K}_h \varphi$ in $D_h$ with $p_i \in a_h(f_h(l))$ every $i = 1, \ldots, k$. In case (a) we have by clause (1) of soundness that $(E, r', m') \models \varphi$. In case (b) we have by clause (2) of soundness that $E \models (p_1 \wedge p_2 \wedge \ldots \wedge p_k) \Rightarrow \varphi$. Since we also have by clause (1) that $(E, r', m') \models p_i$ for every $i$, it follows that $(E, r', m') \models p_1 \wedge p_2 \wedge \ldots \wedge p_k$, and thus $(E, r', m') \models \varphi$, as desired. **QED**

Since $(r, m) \sim_h (r, m)$ for all points $(r, m)$, by definition, we have that standard (possible-worlds) knowledge satisfies the Knowledge Axiom (or knowledge property) $E \models K_h \varphi \Rightarrow \varphi$. Given Theorem 1, it immediately follows that so does human agent knowledge:

**Corollary 1:** If $E$ is sound, then $E \models \hat{K}_h \varphi \Rightarrow \varphi$ holds for all $\varphi \in \mathcal{L}^+$.

Corollary 1 says that when a system $E$ is sound, then every formula $\varphi$ in $\mathcal{L}^+$ that the human agent boundedly knows to be true according to E is indeed true. A second corollary consists of the equivalent, contrapositive, form of Theorem 1:

**Corollary 2**: If $E$ is sound then $E \models \neg K_h \varphi \Rightarrow \neg \hat{K}_h \varphi$.

In other words, if human agent $h$ does not know a proposition $\varphi$ in a system $(R, \pi)$ according to the possible worlds definition of knowledge, then $h$ will not know $\varphi$ in any sound epistemic system $E$ extending the system $(R, \pi)$.

The relationship between knowledge in an epistemic system and the classical possible worlds interpretation of knowledge demonstrates that our approach is fundamentally grounded, with Theorem 1 and its corollaries collectively providing the formal support.

Assume, for instance, that there is an epistemic system $E$ that models the knowledge of the agents in a complex human-automation system $S$. In $E$ there is a local state $l$ of the human agent at which $h$ boundedly knows a proposition $q$. According to Theorem 1, if $E$ is sound, then $h$ must also know $q$ when at $l$ according to the possible worlds approach. Thus if $E$ is sound, any proposition that the human agent knows in our modified approach is indeed true.

This relationship between the two approaches allows us also to demonstrate when the design of a given system $R$ makes satisfactory knowledge impossible. If for example we show that for a state in $R$ satisfying $p_{bad}$ the human agent does not know $p_{bad}$ at that state, then by Corollary 2 no epistemic system for $R$ can be adequate and sound, and the design of $R$ is fundamentally flawed. The design must be changed - perhaps by adding explicit information to the human agent states - before an adequate epistemic setting can be obtained.

Epistemic systems are a framework for reasoning about knowledge in complex human-automation systems that allows us to expressively describe the knowledge a human agent has available for supervisory control. At the same time, the framework allows us to formally determine if that knowledge enables effective system performance.

In the remainder of the paper we discuss a representative case study in which our approach is used to model and analyze the classic *Therac-25 problem* [10]. We use the formalism to describe the knowledge available to the human operator as a function of the system's design, to prove that the existing design precludes effective supervisory control, and to identify the knowledge that is lacking. Following, we show that once missing knowledge is made available (by a change in the interface in this case), effective supervisory control is enabled. In this manner we establish the viability, value, and correctness of thinking formally about human-automation systems at the level of the knowledge in the system, and what agents know.

## V. THE THERAC-25 DEVICE PROBLEM – OVERVIEW

The case study is presented at two levels of detail. The first, immediately following, is a high-level overview discussion that aims to enhance an intuitive understanding of our approach and its value for reasoning about complex human-automation systems. The second, included in the Appendix, is a complete formal representation of the problem and the solution.

### A. Problem Description

"Between June 1985 and January 1987, a computer-controlled radiation therapy machine, called the Therac-25, massively overdosed six people. These accidents have been described as the worst in the [then] 35-year history of medical accelerators" [10].

During investigation of the Therac-25 failures, a common factor in two of the accidents was the device's operator. In both cases, the operator had started and completed the patient data entry task, and then had gone back to edit one or more values before starting the treatment.

Analysis showed that the operator's data entry speed was the cause of the machine's behavior – while editing data was an allowable function, the operator was able to edit the data and return to a 'data entry complete' state so quickly that the machine did not record, and hence was never aware, that the data editing had occurred. The operator then initiated treatment believing that dosage would be in accordance with the newly edited parameter values displayed on the interface, while the actual dosage was given to the patient according to the original data values.

Several flaws in the system's design were implicated in the accidents. The faulty design introduced a system state during which the treatment values could be edited, and the machine activated, without the new values actually being processed. Furthermore, the interface provided almost no information to the operator about the internal behavior of the machine, and the operator thus had no way of knowing whether the machine had accepted her data editing.

*B. Modeling the System*

In order to reason more clearly about the problem and potential solutions we first need to generate a more formal representation of the significant elements of the Therac-25 problem.

*1) Agents*

There are two agents that must be included in the model, the human operator $h$, and the Therac-25 device, denoted $a$. (Reference to the environment is omitted in this example, as the device is deterministic and so the environment plays no essential role.)

*2) Local states*

The human operator's local states in this example capture the information that is available via the Therac-25 device's interface. For instance, the operator's local states will include a variable representing the status of data entry or modification, and a variable for the 'ready-to-treat' status of the device, as presented to the operator by the interface. The machine's local states consist of variables indicating whether there is complete data entry, whether the entered data has been modified, and whether the system is in treat mode.

*3) Actions and joint actions*

In each state, the operator $h$ and the machine $a$ perform actions that may change the global state. The operator of the Therac-25 device can input data, modify the data entered, initiate treatment, or do nothing. The device can process data that has been entered, treat, or do nothing.

*4) Dynamic system behavior*

We represent the dynamic behavior of the combined operator-device system by defining the possible transitions between global states as a function of the operator's and the device's joint actions. For example, consider the global state $g_0$ as the initial state in which the operator has not yet entered any treatment data. A possible joint action in this state would be for the operator to enter treatment data and the Therac device to do nothing. This joint action would cause the system to transition to a global state $g_1$ in which the treatment data has been entered but not yet processed. By defining all possible transitions between states we can obtain a complete model of the system's set of possible behaviors (which will form our set of possible runs $R$). A crucial task in modeling the behavior of the system is to identify behaviors that are unacceptably anomalous, so that the set of points $(r,m)$ in $R$ at which $p_{bad}$ is true may be clearly defined.

*5) Modeling the human's knowledge*

The next step in development of the model is to define and formally represent the operator's knowledge in each of the states of the system; this is the knowledge, the epistemic setup $\Theta_h$, that the operator has in order to control the system.

We start by generating the set of primitive propositions or facts that are relevant to the complete system's dynamic behavior. For instance, facts such as 'no treatment data is entered' and 'device is ready to treat' would be part of this set. The truth value of these facts at a given time will be a function of the actions of the operator and the machine, and the physical dynamics of the complete system. Thus, for example, the proposition 'no treatment data is entered' will be true only if the operator has not yet entered any treatment data.

We then add new propositions of ground or automatic knowledge such as 'data is entered accurately'.

As the final step in modeling the operator's bounded knowledge we define the implications that would comprise the operator's mental model knowledge used to interpret the explicit knowledge in her local state. The Therac operator's mental model might include the implication 'if data entry is complete and the device is ready to treat and the entered data is correct then treatment can be initiated'; in a local state in which the propositions 'data entry is complete', 'the device is ready to treat', and 'entered data is correct' are true, the operator would know that treatment can be initiated. Critically, the operator's mental model must also include the implications that enable knowledge of anomalous system behaviors, so that the agent knows when $p_{bad}$ is true.

## C. Satisfactory Knowledge and Possible-Worlds Knowledge in the Therac-25 Problem

Once the system's behavior and the operator's bounded knowledge have been modeled, we use our definitions of soundness and adequacy to analyze the system and demonstrate that the Therac-25 operator's bounded knowledge as modeled is not satisfactory for supervisory control. Examining the model within the possible worlds framework shows further that the design of the Therac-25 is flawed and so renders satisfactory knowledge impossible.

To begin, recall that the operator of the Therac device boundedly knows that treatment data values can be modified prior to initiating the patient's treatment. We assume that in a given run $r$ the operator modifies previously entered data, the device processes the data and displays its 'ready to treat' status, and the operator initiates treatment. In this run $r$ the device functions as expected and treats the patient in accordance with the modified data; the operator knows the system behavior is acceptable.

Now we consider another run *r'* in which the operator's actions are the same as in run *r* but the device does not process the new data prior to treatment being initiated. Thus $p_{bad}$ holds once treatment is initiated. Since the operator's local states in *r'* are the same as in *r* she will have the same knowledge as in *r*, and so boundedly knows that the system behavior is acceptable. There is no indication in her local states that once treatment is initiated, $p_{bad}$ holds.

- In both runs *r* and *r'*, at the point *m* at which treatment data has been edited, the proposition *p* = 'device is ready to treat' is part of the operator's explicit knowledge, $(E,r,m) \models \hat{K}_h\, p$ and $(E,r',m') \models \hat{K}_h\, p$ . However, since in run *r'* at *(r',m)* the edited data has not yet been processed by the device then *p* is false in *(r',m')*, as is the operator's knowledge of *p*. Soundness, a necessary property of satisfactory knowledge, therefore does not hold.

- In run *r'*, though $p_{bad}$ holds after initiation of treatment, $p_{bad}$ is not in the operator's local state and so adequacy does not hold.

- While at point *(r,m)* treatment data has been edited and processed by the device, there is another point *(r',m')* at which the human operator has the same state as at *(r,m),* so that *(r,m)~$_h$(r',m')*, but at which the new data has not been processed by the device. Thus, while the operator boundedly knows *p* at both *(r,m)* and at *(r',m')*, *p* is true at one point and false at another. Since *(r,m)~$_h$(r',m')*, $K_h p$ (stated in terms of possible-world knowledge) does **not** hold. Corollary 2 implies that no epistemic system for this device can be adequate: there is guaranteed to be a bad state that the operator will be unable to detect as such.

The impossibility of satisfactory knowledge for supervisory control of the Therac-25 system is thus shown. This is discussed further and formally proved in the Appendix.

*D. Problem Resolution*

In addition to supporting a fine-grained analysis of a complex human-automation system, our approach provides clear direction for problem resolution. In the case of the Therac-25 problem, the knowledge-based model and analysis suggests the need for a design modification that provides the operator with the information needed to truly know that the device has processed the operator's edited data values and so is indeed ready to treat.[4] This modification would allow the operator to easily determine the true 'ready to treat' status of the device and so should preclude the possibility of initiating treatment with incorrect treatment data.

*E.  Supporting Design Modification Via Bounded Knowledge*

It remains to demonstrate, using our formal framework, that this modification does indeed result in a system in which the knowledge available is satisfactory for supervisory control. Again assume that there is a state *g* in which treatment data has been edited and processed by the device, and a state *g'* in which treatment data has been edited but not processed by the device.

Let $E^m$ be the epistemic system that models the knowledge of the operator and device in the Therac-25 system modified as described in section 5.4. In the local state *l*, the operator's epistemic setup $\Theta_h(l) = \boldsymbol{d}_h(\boldsymbol{a}_h(f_h(l)))$ now contains a knowledge formula $\hat{K}_h \varphi$ representing the operator's bounded knowledge of the global system's readiness to treat.

For this system $E^m$ to be sound, i.e., for what the operator boundedly knows to indeed be true, the operator must know that the system is ready to treat within the possible worlds framework, as well. It is easy to see that this is so. In any global state *g'* in which treatment data had been edited but not yet processed by the device, the modified device would provide an indication to the operator that the system is not ready to treat. The operator would thus know that the system is ready to treat only in states in which the system truly is ready to treat. That is, in $E^m$ the proposition 'ready to treat' is true in any state or world in which the operator boundedly knows it to be true. This demonstrates that the suggested modification generates a sound epistemic system $E^m$.

Theorem 4 in the appendix states this in more formal terms, and provides proof that the knowledge available with this modification is satisfactory for supervisory control.

*F.  Summary*

The goal of this section was to provide an initial intuitive example of how our approach may be applied to a real system. Though for most complex systems of interest neither the model nor the solution will be nearly as easy to define, this Therac-25 example shows how a 'real-life' system may be modeled, and its human-automation interface design evaluated, using the knowledge of its operator. For the interested reader the complete detailed model and analysis is included in the Appendix.

---

[4] We do not suggest that adding display elements for each fact the human needs to know is a desirable or even viable solution option. Display optimization for the knowledge that is identified as necessary is (far) outside the scope of this research.

## VI.   DISCUSSION AND DIRECTIONS FOR FUTURE RESEARCH

In this article we presented an initial attempt to utilize a well-established formal theory of knowledge and action in multi-agent systems in order to conduct a knowledge-based analysis of a complex human-automation system. This approach allows us to reason cleanly and rigorously about the design and performance of a system as a function of one of its most fundamental resources – the knowledge of the agents. The Therac-25 analysis revealed the existence of design flaws in the system that precluded the human agent from being an effective supervisory controller, and identified the knowledge that was missing. No additional theories of performance or behavior were required, and the human and automation agents were depicted using an expressive common construct without losing important and unique properties of either of these distinct agents. The example demonstrated that our approach offers a formal, expressive, and parsimonious methodology for the design and analysis of complex systems, and we believe this to be a significant contribution. As an initial step in a new direction, the work discussed here raises many intriguing points for investigation; we briefly discuss several that we believe particularly worthy of further pursuit.

### A.  A Richer Notion of Human Knowledge

A first A first direction for further research is to expand on and more completely define a notion of human knowledge within our formal framework. For instance, we may wish to consider a classification of human knowledge in terms of its contents in addition to the type-based taxonomy defined in the current paper, such as the declarative / procedural dimension often used in cognitive modeling [34]-[36]. In this manner we may gain a more multi-dimensional model of human knowledge that would render our approach useful in a wider range of problem domains.

Another aspect of human knowledge that is important in supervisory control is the distinction between knowledge and belief. By capturing the important distinction between a human operator formally knowing a fact $p$ and believing $p$, we can more precisely investigate the impact of incorrect human belief on system performance. Belief has been formally represented using a number of techniques [30], [37], and a valuable goal is to identify and build on the technique most appropriate for modeling a notion of human belief in domains of interest.

An additional element of human knowledge and reasoning that is particularly important in the supervisory control context is the concept of counterfactual reasoning ('if $p$ were to hold then $q$ would be true') [38]. When an operator plans future actions, especially error recovery actions, counterfactual reasoning supports the operator's consideration of conditional alternatives and the outcomes of hypothetical scenarios [39], [40]. The ability to reason about and identify the knowledge needed for effective counterfactual reasoning will be useful in the design of more robust systems that provide the information needed for an operator to successfully 'think through' novel, perhaps safety-critical situations. Preliminary work suggests that incorporating an existing formalization of counterfactual reasoning [27] into our approach will provide designers with a means to do so.

Once these elements of human knowledge are added to our framework, the approach will support the modeling and analysis of a wider and more realistic set of complex systems. For example, a designer could more accurately evaluate the potential failure conditions of supervisory control inherent in a system that executes in a highly ambiguous environment by limiting the human agent's knowledge of important automation and environment agent behaviors. We may be able to capture differences in system performance that result from differences in the amount or quality of knowledge possessed by the human controller. If qualitative differences in knowledge distinguish between novice and expert human operators [41], [42], for example, this would allow a designer to gauge the vulnerability of the system to novice supervisory control and might make salient required emphases in training.

There are, no doubt, many additional aspects of human agent knowledge that are relevant to our problem domain and that are amenable to formal representation in our framework; it is an interesting extension of our work to identify them. Again, the goal should not be to draw a true and faithful picture of human knowledge, but rather to develop a representation that is *epistemically adequate* [43] and that allows us to reason expressively and formally about important properties of human knowledge within the context of complex systems.

## B. Extending the Formal Model

Our formalization of human knowledge has drawn on various approaches in epistemic logic to create a framework that is expressive enough to begin capturing the unique properties of human agent knowledge without sacrificing the rigor of formal logic. Our approach can, as well, formally represent different types of reasoning that a human agent might do, perhaps in different circumstances or in different systems. Previously we noted that our current definition of a 'one-shot' round of reasoning captures an intuitive notion of how a human agent might reason in a supervisory control setting. That is, since complex systems are normally dynamic systems in which control decisions need to be made and implemented in a timely manner, the operator's reasoning activities cannot require multiple rounds of reasoning. One-shot reasoning captures the resource bounds that would be expected due both to the human's inherently limited reasoning abilities as well as to the requirements of a context in which control decisions need to be made and implemented in a timely manner.

It will be interesting to expand on this initial work and identify additional patterns of inference that can be used to express human reasoning, the systems in which they may be appropriate, and the means by which they should be formalized. More precise bounds on reasoning may also be identified. For instance, within the one-shot model, how many independent instances of simple deductive inference can a human do? A system's design might require a human to infer, in the same round, $n$ separate conclusions. Can this requirement be satisfied by typical human cognitive resources? Can we augment the one-shot model with a multi-step (i.e. algorithmic) representation of more complex human reasoning? What might be the natural resource bounds for this type of reasoning actions?

*C. Knowledge of Groups of Agents in Human-Automation Systems*

One of the most significant contributions of the knowledge formalism has been its ability to express notions of the knowledge of groups of agents, such as agents' knowledge of other agents' knowledge, distributed knowledge (knowledge is distributed between the agents in the system), and common knowledge (that is, all the agents know a fact $p$, and know that all agents know $p$, and so on…) [3]. This expressiveness supports analysis of centrally important system properties such as the need for an agent to know what another agent knows, the additional knowledge made available by a fact being commonly known, the potential performance cost when system failure precludes a needed fact from being common knowledge, and so forth.

The importance of being able to reason formally about the knowledge of groups of agents in the design and analysis of human-automation systems is significant, as noted in [44]-[46]. For a simple example, recall the search-and-rescue robot team mentioned previously. Not only must the designer consider what the human and robot agents need to know in any state of the system, but since the robot is normally remotely operated (e.g. it is deep in a collapsed building looking for survivors) the designer must be able to capture what the human and robot agents can know about each other's knowledge at any point in time in order to determine whether the knowledge of the agents will be sufficient for task performance.

It is important to explore the knowledge of groups of agents within the context of human-automation systems. Our work will consider these notions both from a conceptual perspective – for instance, what does it really mean to say that human and automation agents have common knowledge of a fact? – and in terms of formal modeling considerations.

What are the theoretical issues of shared knowledge relevant to a human-automation system? It is natural to say that an operator may need to know what the automation knows, but when is it useful (and meaningful) to talk about an automation agent knowing what a human agent knows or what other automation agents know? Considering various forms of human-robot teams, for example, the need for a robot to know what the human agent knows seems clear in the case of search and rescue, personal assistant, or physical therapy robots [47]. Can we define general taxonomies of human-automation and human-robot systems in which specific types of group knowledge are required for system performance?

The knowledge formalism provides a complete formal semantics and structures for the modeling and analysis of various types of group knowledge. We thus have the tools to represent and reason about the knowledge of *every* agent in the group, $E_G$, *distributed* knowledge in the group, $D_G$, and *common* knowledge, $C_G$.,. An important direction for our work is to further investigate the notion of group knowledge for human-automation systems. As systems grow in size (number of agents), in heterogeneity (types of agents), and in the criticality of the mission, a common concept that binds all agents and that provides a rigorous method for evaluation will become an imperative. We believe this last direction for our research program to be particularly significant.

*D. Applications*

The final test of any formal approach for modeling and analysis is its applicability to real-world problems in the intended domains. The Therac-25 device problem discussed in this article served as a benchmark for demonstrating that our approach (1) can expressively capture important properties of human and non-human agents, (2) can enable us to answer queries about what human and non-human agents know and don't know, and (3) can be used to draw significant conclusions regarding a complex system's design in spite of the dissimilarity of its agents. While these results offer an important initial 'proof-of-concept', the Therac-25 problem is a cleanly bounded and thoroughly researched scenario with faults in design identified *a priori*. What is the value of our formalism as a modeling and analysis tool for systems that are not as neatly defined or for which answers are not as clear?

The state-of-the-art in human-automation system design will provide a valuable test-bed of systems for evaluating our approach 'in real-life'; we are particularly interested in systems in the aviation and human-robot teams domains. In the aviation domain, current directions in integrated flight deck systems design, aviation information management, and the evolving role of the pilot as supervisory controller underscore the inherent relevance of a knowledge-based approach. Will our formalism be able to provide useful insight regarding the design of these highly complex and sophisticated systems?

The application of our formalism to the modeling and analysis of human-robot teams will serve as another challenging and rigorous test of its viability. On the one hand, the importance of capturing notions of agent knowledge and common knowledge, and the need for a formal method for reasoning about knowledge in this domain, have already been noted in the literature [45], [48], and so the potential value of our approach is clear. On the other hand, these systems have unique properties that make accurate and useful representation exceptionally difficult. A knowledge-based model of a human-robot team will need to capture the (albeit artificial) intelligence of the non-human robotic agents and the often hostile environment within which these systems operate (e.g. the collapsed buildings environment of search and rescue teams). Too, our approach must be able to represent and reason usefully about the complex and dynamic patterns of communication and knowledge distribution between multiple human and non-human agents that are to be found in this domain. Consider that in a team of humans and unmanned aerial vehicles (UAVs) the multiple UAVs may communicate among themselves to maintain formation and ensure surveillance coverage while information regarding potential targets is transmitted to the human operator. Or the operator may communicate different commands to different UAVs, and receive different responses, which may be a function of the UAV's reasoning ability and knowledge, rather than just data. How best to capture the role of the knowledge of all the agents in this and similar systems? What insights can we gain here using our knowledge-based approach?

## VII. CONCLUSION

As noted in the introduction to this work, the critical nature of many human-automation systems impose a clear need for rigorous design and analysis tools that are practically useful. This is well-recognized, and the development of methods and tools has been an active area of research for many years. Unfortunately the highly complex nature of many of the tools too often results in their isolation in academic and scientific arenas. Subsequently, practical system design remains to a large extent a somewhat ad hoc process.

Towards that end we have introduced and described the initial development of a novel approach to modeling and reasoning about these systems that is intended to both satisfy the need for formal rigor and be sufficiently intuitive for practical use. One of the most significant aspects of this framework is that agent knowledge is ascribed and analyzed with respect to the automaton representing the complete given human-automation system. Our initial results suggest that reasoning about these systems from the perspective of agent knowledge is in fact a viable and valuable approach. conclusion section is not required. Although a conclusion may review the main points of the paper, do not replicate the abstract as the conclusion. A conclusion might elaborate on the importance of the work or suggest applications and extensions.

APPENDIX

VIII.  THE THERAC-25 DEVICE PROBLEM – DETAILED ANALYSIS

A.  *Problem Description*

"Between June 1985 and January 1987, a computer-controlled radiation therapy machine, called the Therac-25, massively overdosed six people. These accidents have been described as the worst in the [then] 35-year history of medical accelerators" [10].

During investigation of the Therac-25 failures, a common factor in two of the accidents was the device's operator. In both cases, the operator had started and completed the patient data entry task, and then had gone back to edit one or more values before starting the treatment.

Analysis showed that the operator's data entry speed was the cause of the machine's behavior – while editing data was an allowable function, the operator was able to edit the data and return to a 'data entry complete' state so quickly that the machine was never aware that the data editing had occurred. The operator then initiated treatment believing that dosage would be in accordance with the newly edited parameter values displayed on the interface, while the actual dosage given to the patient was as defined by the original data values.

Several flaws in the system's design were implicated in the accidents. The faulty design introduced a system state during which the treatment values could be edited, and the machine activated, without the new values actually being processed. Further, the interface provided almost no information to the operator about the internal behavior of the machine, and the operator thus had no way of knowing whether the machine had accepted her data editing.

*B. Modeling the System*

*1) Agents*

The two agents in our model of the system are the human operator, denoted $h$, and the Therac-25 device, denoted $a$. (Including the environment in this model does not alter the analysis and so we do not add it here.)

*2) Local states*

The local states of each agent consist of the variables or information the agent has available in that state. The human operator's local states, which capture the information available via the interface, include a variable representing the status of data entry or modification, and a variable for the 'ready-to-treat' status of the device. $L_h$ is the set of possible states for the human agent which have the form $l_h = (\mathsf{data.entry, system.status})$, where the possible values for each variable are:

$\mathsf{data.entry} \in \{\varnothing, \mathsf{data.in}, \mathsf{new.data.in}\}$, corresponding to no data entered (i.e. treatment set-up has not yet begun), initial data entry complete, modified data entry complete, respectively. Note that by design the status of data entry is complete only when all required fields are filled and the cursor is in a specified field. This signifies to the system that data entry is complete, and normally triggers the system's processing of the entered data.

$\mathsf{system.status} \in \{\mathsf{sys.ready.no, sys.ready.yes, treat}\}$, indicating whether the system is not ready to treat, ready to treat, or treating.

The machine's local states consist of variables indicating whether there is complete data entry, whether the entered data has been modified, and whether the system is treating. $L_a$ is the set of possible states for the automation device which have the form $l_a = (\mathsf{data, new.data, status})$; the possible values for each variable are

$\mathsf{data} \in \{\varnothing, \mathsf{treat.data}\}$,

$\mathsf{new.data} \in \{\mathsf{new.no, new.yes}\}$,

$\mathsf{status} \in \{\mathsf{ready.no, ready.yes, treat}\}$.

*3) Actions and joint actions*

The operator of the Therac-25 device can input data, modify the data entered, initiate treatment, or do nothing (a null action, denoted $\Lambda$); $\mathrm{ACT}_h = \{input.data, modify.data, press.treat, \Lambda\}$.

The device can process data, treat, or do nothing; $\mathrm{ACT}_a = \{process.data, treat, \Lambda\}$.

The joint actions of the human and automation agents are thus:

(input.data, Λ)

(Λ, process.data)

(modify.data, Λ)

(modify.data, process.data)

(press.treat, treat)

(press.treat, Λ)

(Λ,Λ)

*4) Local states and protocols*

The operator's local states $L_h$ and the protocol $ACT_h$ defining the actions allowable in each local state $l_h$ are listed below.

The human operator's local states are:

$L_h$ = {(∅, sys.ready.no), (data.in, sys.ready.no), (data.in, sys.ready.yes), (new.data.in, sys.ready.no), (new.data.in, sys.ready.yes), (data.in, treating), (new.data.in, treating)}

The operator's protocol $ACT_h$ is:

$ACT_h$(∅, sys.ready.no) = {*input.data*} ∪ {Λ}

$ACT_h$(data.in, sys.ready.no) = {*modify.data*} ∪ {Λ}

$ACT_h$(data.in,sys.ready.yes) = {*press.treat*} ∪ {*modify.data*} ∪ {Λ}

$ACT_h$(new.data.in, sys.ready.no) = {*modify.data*} ∪ {Λ}

$ACT_h$(new.data.in, sys.ready.yes) = {*press.treat*} ∪ {*modify.data*} ∪ {Λ}

$ACT_h$(data.in, treating) = Λ

$ACT_h$(new.data.in, treating) = Λ

The Therac-25 device's local states $L_a$ and the protocol $ACT_h$ defining the actions allowable in each local state $l_a$ are listed below.

Therac-25 local states:

$L_a = \{(\varnothing,$ new.no, ready.no), (treat.data, new.no, ready.no), (treat.data, new.no, ready.yes), (treat.data, new.yes, ready.no), (treat.data, new.yes, ready.yes), (treat.data, new.no, treat), (treat.data, new.yes, treat)$\}$

Therac-25 actions $ACT_a$

$ACT_a(\varnothing,$ new.no, ready.no) $= \Lambda$

$ACT_a($treat.data, new.no, ready.no) $= process.data$

$ACT_a($treat.data, new.no, ready.yes) $= \{treat\} \cup \{\Lambda\}$

$ACT_a($treat.data, new.yes, ready.no) $= process.data$

$ACT_a($treat.data, new.yes, ready.yes) $= \{treat\} \cup \{\Lambda\}$

$ACT_a($treat.data, new.no, treat) $= treat$

$ACT_a($treat.data, new.yes, treat) $= treat$

### 5) *Dynamic system behavior*

The system's dynamic behavior is represented as the set of runs *R* that result from the operator's protocol above and the automaton's protocol, in which the transitions from one global state to the next are in accordance with a transition function τ; the complete set of possible runs is defined to be all executions of the automaton in Fig. 1. Note that we assume a correctly operating device and so anomalous behaviors that might arise should the device malfunction are not included. We also bound the scope of the model by defining the states in which treatment has been initiated as end states. Finally, in the interest of readability the automaton does not display the self-loops that result from joint null actions (Λ,Λ) or from actions that have no effect (e.g. the operator action *treat* when the machine is 'not.ready'.)

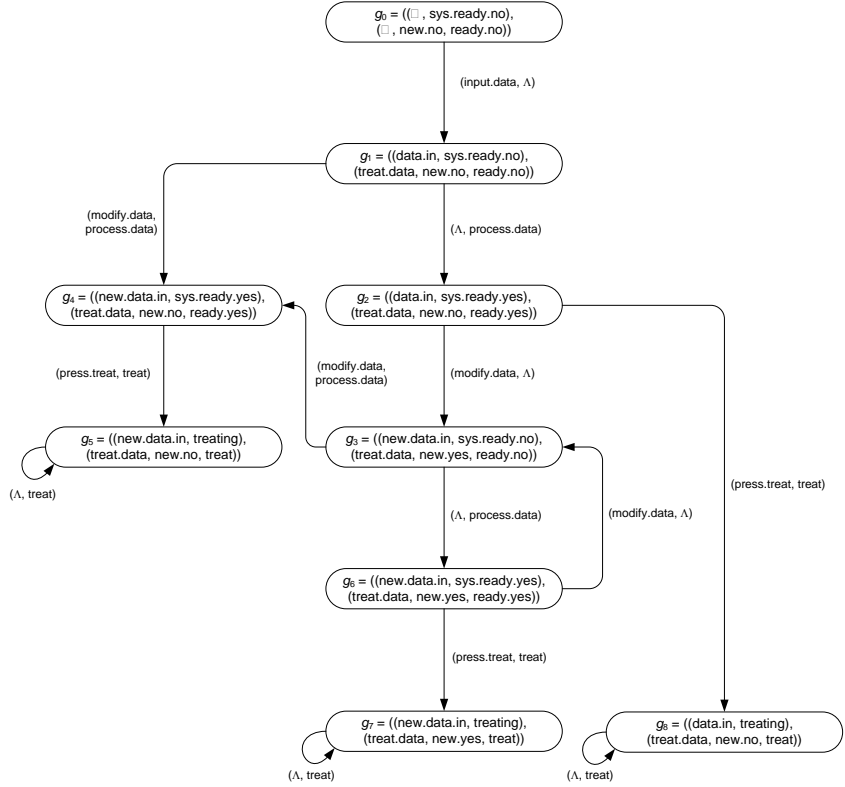

Fig. 1. An automaton describing dynamic behavior of the Therac human-automation system

## C. Impossibility of a Sound Epistemic Setup for the Therac Example

A logical treatment of the Therac example would start defining a set $\Phi$ of primitive propositions or facts that are relevant to the system's dynamic behavior; For example, we may have $\Phi = \{p_1, \ldots, p_6\}$ where

$p_1 =$ 'no treatment data is entered'

$p_2 =$ 'initial data entry is complete'

$p_3 =$ 'modified data entry is complete'

$p_4 =$ 'modified data entry has been processed'

$p_5 =$ 'device is ready to treat'

$p_6 =$ 'device is treating'

$\Phi$ is derived from the Therac 25 system specification; it represents system design assumptions regarding the knowledge an operator is required to have. Recall that, in general, the proposition $p_{bad}$ should capture the statement 'the current state of the system is not acceptable'. In every particular system, $p_{bad}$ will correspond to a property specific to that system. In the Therac-25 example, we identify $p_{bad}$ with $\neg p_4 \wedge p_6$ : A state in which the operator has modified the data and the device is treating but the treatment is not in accordance with the defined values.

At any given time in system behavior, the truth value of each proposition in $\Phi$ is well-defined based on its description in words above. This truth value is a function of the agents' actions and the physical dynamics of the complete system. For instance, the proposition $p_3$ standing for 'modified data entry complete' will be true only if the operator has indeed completed data entry and returned the cursor to the correct field; the proposition $p_4$, 'modified data entry has been processed,' holds only when the Therac device has recognized that the data has been modified and has reset its treatment parameters. More specifically, the interpretation $\pi$ that maps the propositions $p_i$ $\in \Phi$ and the global states $g$ to True or False captures the behavior of the Therac human-automation system as designed.

In this post hoc analysis, we can straightforwardly show that no sound epistemic system exists for the design depicted in Figure 1.

**Theorem 2:** There can be no adequate epistemic setup $E$ for the existing design of the Therac-25

Proof: As suggested in the main text, we prove this result via the connection between possible-worlds knowledge and bounded knowledge. Let $R$ be the set of runs of the automaton depicted in Figure 1. Let $r \in R$ be a run whose first four global states are $g_0$, $g_1$, $g_4$ and $g_5$, and similarly let $r'$ be a run with prefix $g_0$, $g_1$, $g_2$, $g_3$, $g_6$, $g_7$. Observe that $r(4)=g_5$, while $r'(6)=g_7$. In the run $r$ the operator enters initial data, and then enters a modified version of the data, while the device processes the initial data, and signals that data entry is complete. Once the system ready indication is displayed the operator initiates treatment. Treatment is not, however, in accordance with the new data and thus $(R,r,4) \models p_{bad}$. In the run $r'$, the operator enters initial data, this data is processed by the device, the operator enters modified data, and the new data is also processed. In this case, the processed data coincides with the latest entered data, and so $(R,r',6) \models \neg p_{bad}$. The crucial point is that $(r,4) \sim_h (r',6)$. It follows that $(R,r,4) \models p_{bad} \wedge \neg K_h \, p_{bad}$. By Corollary 2, we have that $(E,r,4) \models p_{bad} \wedge \neg \hat{K}_h \, p_{bad}$ for every epistemic setup $E$ for the system $R$. It follows that there can be no adequate epistemic setup for $R$.

*D. Problem Resolution*

Our approach not only identifies problems with agent knowledge, but provides insight into problem resolution and the means to determine whether the resolution is adequate. For the Therac-25 system, the modeling and analysis process suggests possible and provably correct solutions. In the situation described above it was shown that the operator had no way of determining that the device had processed the data, nor was she aware of the bad state once treatment was initiated prior to data processing. A straightforward design solution would be to modify the interface so that information regarding the device's data processing status would be always available. For instance, the operator's modification to the entered data might disable further system action and cause the system ready indicator to turn red. Only once new data had been completely entered and then processed by the device would the device display a data.ready signal to the control panel, in the operator's local view. In this manner, the human operator would be able to know when treatment could be initiated safely.

A sound and adequate epistemic setting for this modified design would have the following features. We add another proposition $p_7$ to $\Phi$, standing for "the value of data.ready is no". The operator's bounded knowledge of system readiness to treat after modification of treatment values would then be based on the explicit information available on the interface display. It is straightforward to see that this change eliminates the design flaw that precluded satisfactory operator knowledge.

*E. Conclusion*

This Therac-25 problem model and its suggested solution were simplified in order to serve as a straightforward example of how our approach may be applied. Note that multiple iterations of the analysis and solution definition process will derive a system design that is provably satisfactory for supervisory control. Though for most complex systems of interest neither the model nor the solution will be nearly as easy to define[5], the Therac-25 example has shown how a 'real-life' system may be modeled, and its human-automation interface design evaluated, using the knowledge of its operator.

---

[5] We also do not suggest that adding display elements for each fact the human needs to know is a desirable or even viable solution option. Display optimization for the knowledge that is identified as necessary is (far) outside the scope of this research.